\documentclass[aps,prl,reprint,singlecolumn,superscriptaddress]{revtex4}

\usepackage{ucs}
\usepackage[utf8x]{inputenc}

\usepackage[english]{babel}
\usepackage{color,graphicx}
\usepackage{amsmath,amsfonts,amssymb,color,fancybox}
\usepackage{units}
\usepackage{mhchem}
\usepackage{epstopdf}

\begin{document}

\preprint{APS/123-QED}

\title{\textcolor{black}{Entropy production selects nonequilibrium states in multistable systems}}


\author{Robert G. Endres}
\affiliation{Department of Life Sciences, Imperial College, London SW7 2AZ, United Kingdom}
\affiliation{Centre for Integrative Systems Biology and Bioinformatics, Imperial College, London SW7 2AZ, United Kingdom}
\email[E-mail: ]{r.endres@imperial.ac.uk}

\date{\today}

\begin{abstract}
Far-from-equilibrium thermodynamics underpins the emergence of life, but how has been a long-outstanding puzzle. Best
candidate theories based on the maximum entropy production principle could not be unequivocally proven, in part due
to complicated physics, unintuitive stochastic thermodynamics, and the existence of alternative theories such as the 
minimum entropy production principle. Here, we use a simple, analytically solvable, one-dimensional bistable chemical system 
to demonstrate the validity of the maximum entropy production principle. 
To generalize to multistable stochastic system, we use the stochastic least-action principle  
to derive the entropy production and its role in the stability of nonequilibrium steady states. 
This shows that in a multistable system, all else being equal, the steady state with the highest entropy production is favored, 
with a number of implications for the evolution of biological, physical, and geological systems.

\end{abstract}


\maketitle

The second law of thermodynamics is often misused to explain that life’s order, e.g. that of DNA, proteins, and cells, 
cannot emerge by chance (or at least is infinitesimally unlikely) \cite{schopf02}. In an isolated system, e.g. the whole universe, the 
entropy (‘disorder’) has to either stay constant or increase. However, in open systems, characterized by fluxes of 
energy and matter, order can arise as long as the entropy of the {surrounding system} increases enough so that the total 
entropy {from the two parts of the system together} increases \cite{schroedinger44}. Note also that the second law does not make 
any predictions about how fast 
a system approaches equilibrium, {except at the stationary nonequilibrium steady state. Under this circumstance, the 
Carnot efficiency limits the rate at which entropy is produced by the heat flux into the system \cite{yoshida08,kawazura12}}.
This qualitative nature of the second law makes the prediction of dynamical systems based on thermodynamics notoriously 
difficult.

For the last 150 years, there has been speculation that universal extremal principles determine what happens in 
nature \cite{helmholtz1968,rayleigh1913,onsager1931ab}, 
most prominent being the maximum entropy production principle (MaxEPP) by Paltridge, Ziegler and others \cite{paltridge78,ziegler77} 
(see also \cite{whitfield05,dewar_book14} for reviews). Its most important conclusion 
is that there is life on Earth, or the biosphere as a whole, because ordered living structures help dissipate the energy from the sun on 
our planet more quickly as heat than just absorption of light by rocks and water \cite{kleidon04,england13,england15}. 
However, such principles have never rigorously been proven, and conflicting results exist. MaxEPP apparently explains 
Rayleigh-Bénard convection, flow regimes in plasma physics, the laminar-turbulent flow transition in pipes, 
crystallization of ice, certain planetary climates, and ecosystems \cite{lorenz01,whitfield05,martyushev07,jesus12}. 
For instance, in plasma physics large-scale dissipative structures can increase the impedance and thus sustain high temperature
gradients while producing large amounts of entropy at a smaller scale \cite{yoshida08}.

Confusingly, MaxEPP seems to contradict the minimum entropy production principle (MinEPP) as promoted by Onsager, Prigogine and
others \cite{onsager1931ab,pregogine67}. MinEPP applies close to equilibrium as an extension of Rayleigh's principle, and predicts the 
current distribution in parallel circuits (Kirchhoff’s law) and the viscous flow in Newtonian fluids \cite{jaynes80}. To make things worse, 
both MinEPP and MaxEPP can apply simultaneously \cite{niven10}, or one of the two can be selected depending on the boundary conditions, 
i.e. whether fluxes (e.g. currents) or forces (e.g. temperature gradients) are constrained \cite{yoshida08,kawazura10,kawazura12} 
(this issue is revisited in the Discussion section). These wide-ranging results, along with the broad range of applications from biochemistry, 
fluid mechanics, ecosystems, and whole planets, leave the question of extremal principles wide open \cite{martyushev10}.

What about more general theoretical approaches? A promising direction for proving the MaxEPP is based on 
information-theoretic approaches related to the maximum-entropy inference method \cite{dewar03,dewar05}, but previous attempts 
relied on overly strong assumptions (reviewed in \textcolor{black}{\cite{bruers07,dewar09,ross12}}). One extremal principle is undisputed – the least-action principle 
(for conservative systems), which can be used to derive most physical theories, including Newton’s laws, Maxwell’s equations, and 
quantum mechanics \cite{feynman06,doi11}. Recently, the stochastic least-action principle was also established for dissipative systems 
\cite{bialek00,wang06,seifert12}. Information theory and the stochastic least-action principle are important corner stones of modern 
stochastic thermodynamics \cite{seifert12}.

Here, with an interest in the emergence of protocells and life, we focus on stochastic biochemical systems
and ask whether MaxEPP provides a mechanism for selecting states in a multistable system. We demonstrate that previous attempts 
to disprove MaxEPP suffered from {misinterpretations of unintuitive aspects} of stochastic systems, and that 
with modern approaches in stochastic thermodynamics, 
MaxEPP can be proven. Initially, we focus on the single chemical species, one-dimensional (1D) bistable Schl\"ogl model, 
but then generalize to stochastic multistable systems. In particular, similar to \cite{meysman10}, we distinguish between two types of MaxEPPs: 
the `state selection' principle, addressing which steady state is selected in a multistable system, and the `gradient response' principle,
focusing on the response of the average entropy production of a stochastic system to changes in a parameter. We find that if
multiple steady states exist, then the first MaxEPP predicts that the steady state with the 
highest entropy production is the most likely to occur. 
Furthermore, we demonstrate that MinEPP simply corresponds to the second MaxEPP near equilibrium. These findings should 
clarify the role of thermodynamics in selecting nonequilibrium states in biological, chemical, and physical systems.\\

\begin{figure*}[t]
\includegraphics[width=10cm]{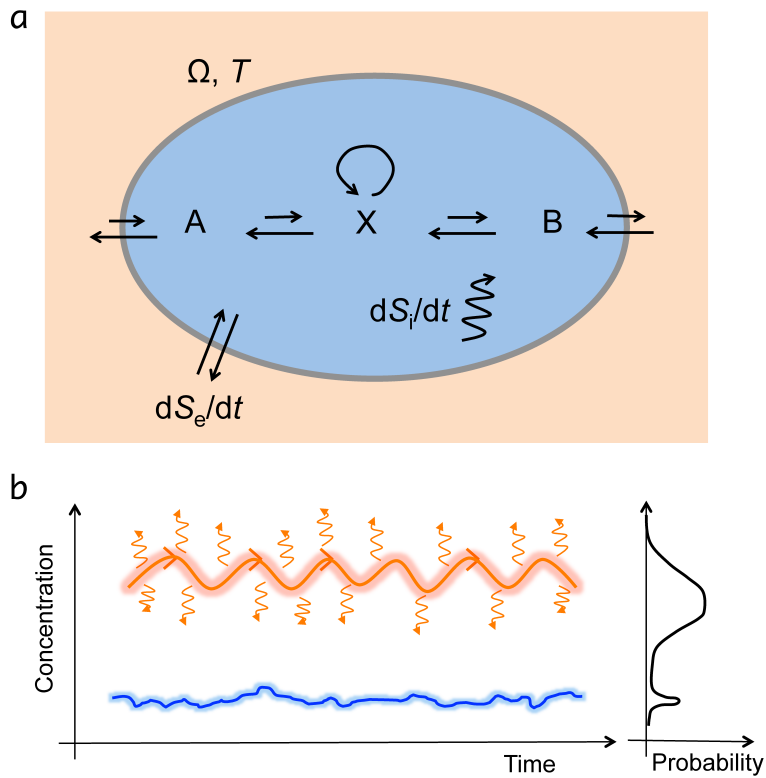}
\caption{{\bf Illustration of a driven system with sources of entropy production and flow.} 
(a) Picture of a driven system (similar to Schl\"ogl model) where fluxes of molecular species $A$ and $B$ drive the concentration of 
species $X$ out of equilibrium. \textcolor{black}{A system is called {\it closed} if only energy is exchanged with the surroundings,
and {\it open} if also matter is exchanged. (For completeness, in an isolated system, there is no exchange at all with the surroundings. 
In a special case of the latter, everything is included, i.e. the subsystem and its surroundings.)} 
Entropy flow rate $dS_e/dt$ across boundary of reaction volume $\Omega$ can have contributions from heat and material flow. Entropy production rate $dS_i/dt$ 
is due to processes inside the volume. At steady state, both \textcolor{black}{time-averaged} contributions are equal in magnitude 
but of opposite sign (see text for details). (b) Illustration of the dynamics of a bistable system with 
a `low' and a `high' molecule-concentration steady state (left). The high state produces high amounts of entropy as indicated by heat radiation, 
warm glow, directedness and oscillatory fluxes (projected {onto the} concentration axis), while the low state is cold and close to 
equilibrium with little entropy production. The hypothesis of MaxEPP is that the state with high entropy production also is 
the more likely to occur (right).}
\label{fig1}
\end{figure*}

\noindent{\bf{A short primer on entropy production}}\\
Imagine a system undergoing state changes due to external driving as shown in \textcolor{blue}{Fig. \ref{fig1}a}. 
If this is done reversibly, the total change in entropy is $\Delta S=\int dQ/T$, where $T$ is the temperature and $Q$ is the added heat. Hence, 
$\int dQ/T$ is called the entropy flux ($\Delta S_e$) from the environment into the system. If done irreversibly, $\Delta S\geq\int dQ/T$. The
positive difference is the entropy production $\Delta S_i$ \cite{landi13}
\begin{equation}
\Delta S - \underbrace{\int\frac{dQ}{T}}_{\Delta S_e}=\Delta S_i\geq 0,
\end{equation}
which after division by $\Delta t$ becomes in the infinitesimal limit
\begin{equation}
\frac{dS}{dt}=\frac{dS_i}{dt}+\frac{dS_e}{dt},
\end{equation}
where $dS_i/dt\geq 0$ is the non-negative entropy production rate and $dS_e/dt$ is the entropy flow rate, which 
can be positive or negative. \textcolor{black}{The entropy flow rate $dS_e/dt=T^{-1}(dQ/dt+\sum_{r}dn_r/dt\,\Delta\mu_r)$ can have contributions 
from heat flow ($dQ/dt$) at temperature $T$ (not part of Schl\"ogl model) and material flow (due to current $dn_r/dt$ and 
and chemical potential difference $\Delta\mu_r$ for reaction $r$).}
However, at a nonequilibrium steady state, \textcolor{black}{the time-averaged rate of entropy
change is zero, i.e. $\overline{dS/dt}=0$, and the entropy production rate
\begin{equation}
\overline{\frac{dS_i^*}{dt}}=-\overline{\frac{dS_e^*}{dt}}\geq 0\label{equal_dSdt}
\end{equation}
is positive in line with the second law of thermodynamics. Put differently, the time-averaged} entropy production rate is the 
negative of the entropy flow rate. Since entropy flow is often easier to calculate, it can be used in place of the entropy production at steady state. 
At thermodynamic equilibrium, both quantities are zero. Now, we need a microscopic model to investigate this further.

Consider a system characterized by a molecular species with molecule number given by $X$ (this can easily be generalized to higher dimensions or
more chemical species), 
where reaction events produce random jumps in molecule number \textcolor{black}{$X\rightarrow X+\Delta X_r$ with transition rate $W_r(X|X+\Delta X_r)$}
(using notation from \cite{gaspard04}).
The backward (time-reversed) reaction $-r$ can also occur, producing random jumps \textcolor{black}{$X+\Delta X_r\rightarrow X$ with transition rate 
$W_{-r}(X+\Delta X_r|X)$.} 
How do we obtain the probability distribution $P(X,t)$ based on these transition rates? The master equation describing the exact 
time evolution of $P(X,t)$ is given by \cite{jiu-li84,lebowitz99,gaspard04}
\textcolor{black}{\begin{equation}
\frac{d}{dt}P(X,t)=\sum_{r}J_r(X,t)\label{master}
\end{equation}
with flux 
\begin{equation}
J_r(X,t)=W_{r}(X-\Delta X_r|X) P(X-\Delta X_r,t) - W_{-r}(X|X-\Delta X_r) P(X,t),\label{J}
\end{equation}}
where the first (second) term on the right-hand side corresponds to an increase (decrease) in probability $P(X,t)$ by jumps
towards (away from) the state with $X$ molecules.
Generally, for nonequilibrium systems detailed balance \textcolor{black}{$W_{r}(X-\Delta X_r|X)P(X-\Delta X_r)=
W_{-r}(X|X+\Delta X_r)P(X)$}
is not fulfilled. Hence, apart from a transient, $P(X,t)$ will relax in time towards a nonequilibrium steady state (from now on simply 
called `state') with time-independent $P(X)$. For a nonequilibrium system, the Gibbs-Boltzmann entropy of statistical mechanics 
for probabilities $P(X,t)$ is given by \cite{jiu-li84,gaspard04}
\begin{equation}
S(t)=\sum_{X}S^0(X)P(X,t)\,-\,\sum_{X}P(X,t)\ln P(X,t)\label{Sdot}
\end{equation}
\textcolor{black}{in units of Boltzmann's constant $k_B$ and $P(X,0)$ the initial condition (prior).} 
The first term on the right-hand side of Eq. \ref{Sdot} describes the average entropy $S^0(X)$ of a 
fixed number of molecules due to internal degrees of freedom, while the second term describes the average entropy due to probability distribution $P(X,t)$
itself. When differentiating with respect to time, we obtain for entropy flow and entropy production rates
\textcolor{black}{\begin{subequations}
\begin{eqnarray}
\frac{dS_e}{dt}&=&\underbrace{\sum_{X,r}S^0(X)J_r(X,t)}_{\rm passive}-
   \underbrace{\frac{1}{2}\sum_{X,r}J_r(X,t)\ln\frac{W_{r}(X-\Delta X_r|X)}{W_{-r}(X|X-\Delta X_r)}}_{\rm active}\label{Sedot}\\
\frac{dS_i}{dt}&=&\frac{1}{2}\sum_{X,r}J_r(X,t)\ln\frac{W_{r}(X-\Delta X_r|X)P(X-\Delta X_r,t)}{W_{-r}(X|X-\Delta X_r)P(X,t)}\geq 0\label{Sidot},
\end{eqnarray}
\end{subequations}}
respectively. The passive contribution in Eq. \ref{Sedot} is due to advection of molecules of a certain complexity (entropy), while the
active contribution is due to changes in number $X$ of molecules.
The nonnegativity of $dS_i/dt$ is due to inequality $(R_+-R_-)\ln(R_+/R_-)\geq 0$ \textcolor{black}{with $R_+$ and $R_-$ generic forward and backward
reaction rates}, and demonstrates the second law of thermodynamics 
\textcolor{black}{(see \textcolor{blue}{Supplementary Information} for further explanations). Note, that the active part of 
Eq. \ref{Sedot} can also be written more compactly \cite{lebowitz99}
\begin{equation}
\frac{dS_e}{dt}=-\sum_{X,r}W_{r}(X-\Delta X_r|X) P(X-\Delta X_r,t)\ln\frac{W_{r}(X-\Delta X_r|X)}{W_{-r}(X|X-\Delta X_r)},\label{Sedot2}
\end{equation}
which again can be used to conveniently calculate the rate of entropy production at steady state upon change of the overall sign.
Furthermore,} Eq. \ref{Sidot} represents a lower bound of the entropy production, assuming infinitely fast mixing \cite{esposito12,ziener15}.

Another way to look at stochastic systems and entropy production is through trajectories $\Gamma$, defined by temporally ordered 
numbers of molecules $\Gamma(t)=X_0\rightarrow X_1\rightarrow X_2\rightarrow\dots\rightarrow X_n$ for times $0<t_1<t_2<\dots<t_n<t$. The
backward trajectory is then given by $-\Gamma(t)=X_n\rightarrow\dots\rightarrow X_2\rightarrow X_1\rightarrow X_0$. According to the
Evans-Searles fluctuation theorem \cite{seifert12}, the entropy change along the trajectory $\Gamma$ is given by the log-ratio of their 
individual constituent probabilities
\textcolor{black}{\begin{equation}
\Delta S_\Gamma=\ln\frac{P(X_0)}{P(X_n)}+\ln\frac{P_\Gamma}{P_{-\Gamma}}\approx\ln\frac{W_{r1}(X_0|X_1)W_{r2}(X_1|X_2)\dots W_{rn}(X_{n-1}|X_n)}
{W_{-r1}(X_1|X_0)W_{-r2}(X_2|X_1)\dots W_{-rn}(X_{n}|X_{n-1})},\label{FT}
\end{equation}
where we neglected $\ln\frac{P(X_0)}{P(X_n)}$ with prior $P(X_0)$ on the right-hand side, valid for long trajectories.} This final expression 
is sometimes called the medium entropy \cite{seifert12} or action functional \cite{lebowitz99,gaspard04}
but in our terminology corresponds to the entropy flow. Equation \ref{FT} reflects our intuition that
the more the entropy increases during the forward process (along $\Gamma$), the less likely is the backward process ($-\Gamma$), 
reflecting the breaking of time-reversal symmetry.
(However, this also shows that small stochastic systems can violate the second law of thermodynamics!) At steady state, we again obtain the 
entropy production, and the ensemble averaged $\langle\Delta S_\Gamma(t)\rangle$, averaged over trajectories $\Gamma(t)$ of duration $t$, 
is given by the time-integrated entropy production rate \cite{gaspard04}
\textcolor{black}{\begin{equation}
\langle\Delta S_\Gamma(t)\rangle=\int_0^t\frac{dS_{i}^*}{d\tau}d\tau=t\overline{\frac{dS_i^*}{d\tau}},
\end{equation}
valid in the long-time limit, and where we defined $\langle\Delta S_\Gamma(t)\rangle=\sum_{\Gamma}P_{\Gamma}S_{\Gamma}(t)$}. 
In the following, we use both the molecule number and trajectory-based pictures.
\ \\
\ \\
\noindent{\bf Minimal nonequilibrium bistable model}\\
To understand the validity of the MaxEPP, why not investigate it with a simple exactly solvable model? This was indeed attempted using 
the well-known chemical Schl\"ogl model {(of the second kind) \cite{schloegl72,vellela09}}. The results were used as an 
argument against MaxEPP, but, as we highlight later, there are issues with this argument due to Keizer's paradox.
\textcolor{black}{This paradox highlights the fact that microscopic (master equation) and macroscopic mean-field descriptions can yield 
very different results \cite{kurtz71,kurtz72,vellela07}. Here, we describe in more detail the Sch\"ogl model, followed by an
explanation of Keizer's paradox in the subsequent section.}

The Schl\"ogl model only depends on one chemical species $X$ with interesting features such as bistability (two different stable 
steady states), first-order phase transition (energy-assisted jumps between states), and front propagation in spatially 
extended systems \cite{vellela09,endres15}. Biochemically, the model converts species $A$ to $B$ and vice versa via intermediate species $X$ 
\begin{subequations}
\begin{eqnarray}
A &\underset{k_{-1}}{\stackrel{k_1}{\rightleftharpoons}}& X \label{A}\\
3X &\underset{k_{-2}}{\stackrel{k_2}{\rightleftharpoons}}& 2X\,+\,B,\label{B}
\end{eqnarray}
\end{subequations}
with rate constants as shown (note we use same capital letter symbols for species names and molecule numbers.)
The model recently attracted renewed interest due to its mapping onto biologically relevant models with bistability, e.g. \cite{scheffer01,ozbudak04,endres15}, 
with the caveat that Michaelis-Menten enzyme kinetics need to be replaced by mass-action kinetics \cite{endres15} or appropriately Taylor expanded 
(see \textcolor{blue}{Supplementary Information} for an example and \cite{wilhelm09} for other minimal biological models based on 
mass-action kinetics.) Nevertheless, the Schl\"ogl model (with spatial dependence and diffusion) is believed to describe front propagation in 
CO oxidation on Pt single crystals surfaces, and the nonlinear generation and recombination processes in semiconductors \cite{ertl15}.

In terms of the master equation, the transition rates are  \cite{gaspard04,vellela09,endres15}
\textcolor{black}{\begin{subequations}
\begin{eqnarray}
W_{+1}(X|X+1)&=&k_{+1}A\\
W_{-1}(X+1|X)&=&k_{-1}(X+1)\\  
W_{+2}(X|X-1)&=&k_{+2}X(X-1)(X-2)/\Omega^2\\
W_{-2}(X-1|X)&=&k_{-2}B(X-1)(X-2)/\Omega,
\end{eqnarray}
\end{subequations}}
{where the molecule numbers of species $A$ and $B$ are fixed. 
Such a chemical system can be simulated by the Gillespie algorithm, which is a dynamic Monte Carlo method and reproduces the exact 
probability distribution from the master equation upon long enough simulations (sampling)
(\textcolor{blue}{Fig. \ref{fig2}a,b}) \cite{gillespie77}.

For large (but finite) volumes $\Omega$, an analytical formula for the probability distribution can be derived, i.e. 
$p(x)=N(x)\exp[-\Omega\Phi(x)]$, with lengthy expressions for normalization $N(x)$ and nonequilibrium potential $\Phi(x)$  
(see \textcolor{blue}{Supplementary Information} for details on the large-$\Omega$ limit of the master equation \cite{hanggi84}; 
\textcolor{blue}{Supplementary Fig. S1}  compares this potential with other potentials used later.) 
Note, while the prefactor $N(x)$ has a weak $x$-dependence, the main $x$-dependence comes from the
potential in the exponential, which dominates for large $\Omega$ ($p(x)=\exp[-\Omega\Phi(x)+\ln N(x)]\approx N(x^*)\exp[-\Omega\Phi(x)]$, 
where $x^*$ is a steady-state value, which minimizes $\Phi(x)$; see \cite{hanggi84}).
While the expression for $\Phi(x)$ is lengthy, its spatial derivative
\begin{equation}
\Phi'(x)=\ln\left(\frac{w_{+1}+w_{-2}-w_{-1}-w_{+2}}{w_{+1}+w_{-2}+w_{-1}+w_{+2}}\right)\label{phi'}
\end{equation}
is simple (and will be used later), {with $w_{\pm r}=W_{\pm r}/\Omega$ for $r=1,2$.} 

However, for easier analytical calculations, there are a number of possible simplifying assumptions to the master equations.
In the macroscopic (infinite volume) limit, the dynamics of the concentration 
\begin{equation}
x=\frac{1}{\Omega}\sum_{X=0}^\infty X\,P(X,t)\label{x}
\end{equation}
(written in small letter symbols) is described by the deterministic ordinary differential equation (ODE)
\begin{equation}
\frac{dx}{dt}=-\underbrace{k_{+2}x^3}_{w_{+2}}+\underbrace{k_{-2}bx^2}_{w_{-2}}
-\underbrace{k_{-1}x}_{w_{-1}}+\underbrace{k_{+1}a}_{w_{+1}}=-\Phi'_{\rm ODE}(x).\label{ODE}
\end{equation} 
{Since $dx/dt$ resembles the velocity of an overdamped particle at position $x$, the right-hand side of
Eq. \ref{ODE} is written in terms of the gradient of an effective potential $\Phi'_\text{ODE}(x)=d\Phi_\text{ODE}(x)/dx$,
where $\Phi'_{\rm ODE}(x)$ does not have units of energy over length but concentration over time.
Parameters $a=A/\Omega$ and $b=B/\Omega$ describe the concentrations of the externally fixed reservoir species, used for driving the 
system out of equilibrium.} Eq. \ref{ODE} can be used to produce the steady-state ($dx/dt=0$) bifurcation diagram, e.g. as a function of 
$b$ (\textcolor{blue}{Fig. \ref{fig2}c}, black line). This demonstrates bistability in a regime of intermediate concentrations of $b$ 
(two stable steady states and one intermediate unstable state). However, the deterministic model does not predict the weights of the steady 
states, {i.e. the probabilities of the steady states}, which requires solving the corresponding stochastic master equation. 

To quantify the entropy production rate for maintaining the steady state, the limit $\Omega\rightarrow\infty$ of Eq. \ref{Sidot} can be taken,
leading to \cite{gaspard04}
\begin{equation}
\frac{ds_i}{dt}=\sum_{r=1}^2\underbrace{(w_{+r}-w_{-r})}_{\text{flux}}\underbrace{\log\left(\frac{w_{+r}}{w_{-r}}\right)}_{\Delta\mu/T}\geq0,\label{sidot}
\end{equation}
with $ds_i/dt=\Omega^{-1}dS_i/dt$ and units of $k_B$ (where we omitted the asterix from Eq. \ref{equal_dSdt} for indicating steady state). 
In Eq. \ref{sidot}, the sum is over reaction types, and the factor ($w_{+r}-w_{-r}$) and the log term represent 
the flux and the chemical-potential (or Gibbs free-energy) difference (divided by temperature $T$) of each reaction, respectively 
\textcolor{black}{(see \textcolor{blue}{Supplementary Information} for a derivation).} This entropy production is illustrated
in \textcolor{blue}{Fig. \ref{fig1}b}, which shows both a low molecule-number state dissipating little, as well as a high-molecule number state 
dissipating a lot. At equilibrium for $b_0=1/6$
all subreactions fulfill detailed balance, i.e. $w_{+r,0}=w_{-r,0}$ with subscript $0$ indicating equilibrium rates, and the entropy production rate
is zero. As a consequence, near equilibrium MinEPP is valid: expanding $w_{\pm r}=w_{\pm r,0}+\delta w_{\pm r}$, 
using $\delta\Delta w_r=\delta w_{+r}-\delta w_{-r}$ 
and $\ln(1+\delta x)\approx \delta x$ for small $\delta x$, we obtain $ds_i/dt\approx w_{-r,0}^{-1}\delta\Delta w_r^2$, i.e. a quadratic form 
with positive prefactor. Hence, near equilibrium the entropy production rate is minimized {with respect to changes in the
rates (or their parameters).}

It is important to note that Eq. \ref{ODE} is indistinguishable from an equilibrium system, e.g. as the equation is equivalent to
an overdamped particle in an anharmonic potential, and the entropy production could mistakenly be written like 
$ds_i/dt=(w_{+1}+w_{-2}-w_{-1}-w_{+2})\ln[(w_{+1}+w_{-2})/(w_{-1}+w_{+2})]=0$, which may appear like an extreme version 
of MinEPP! To remedy this problem, we can rewrite Eq. \ref{ODE} by moving away from concentration constraints (forces) to flux constraints
\begin{subequations}
\begin{eqnarray}
\frac{dx}{dt}&=&w_{+1}-w_{-1}+w_{-2}-w_{+2}\label{ODEx}\\
\frac{da}{dt}&=&-w_{+1}+w_{-1}+F\label{ODEa}\\
\frac{db}{dt}&=&-w_{-2}+w_{+2}-F\label{ODEb},
\end{eqnarray}
\end{subequations}
where Eqs. \ref{ODEa} and \ref{ODEb} represent explicitly the dynamics of the reservoir species $a$ and $b$. Imposed flux $F$ ensures 
that the molecule concentrations $a$ and $b$ are maintained and that the dynamics of species $x$ are driven out of equilibrium.
At steady state (given by $x^*$, $a^*$, and $b^*$), 
the flux $F=w_{+1}-w_{-1}=w_{+2}-w_{-2}$ produces entropy at a rate given by $ds_i/dt=F[\ln(w_{+1}/w_{-1})+\ln(w_{+2}/w_{-2})]$,
and hence Eq. \ref{sidot} follows naturally as the sum of entropy production rates of the individual reactions, Eqs. \ref{ODEx}-\ref{ODEb}.
The entropy production rate is plotted in \textcolor{blue}{Fig. \ref{fig2}d}.

\begin{figure*}[t]
\includegraphics[width=15.5cm]{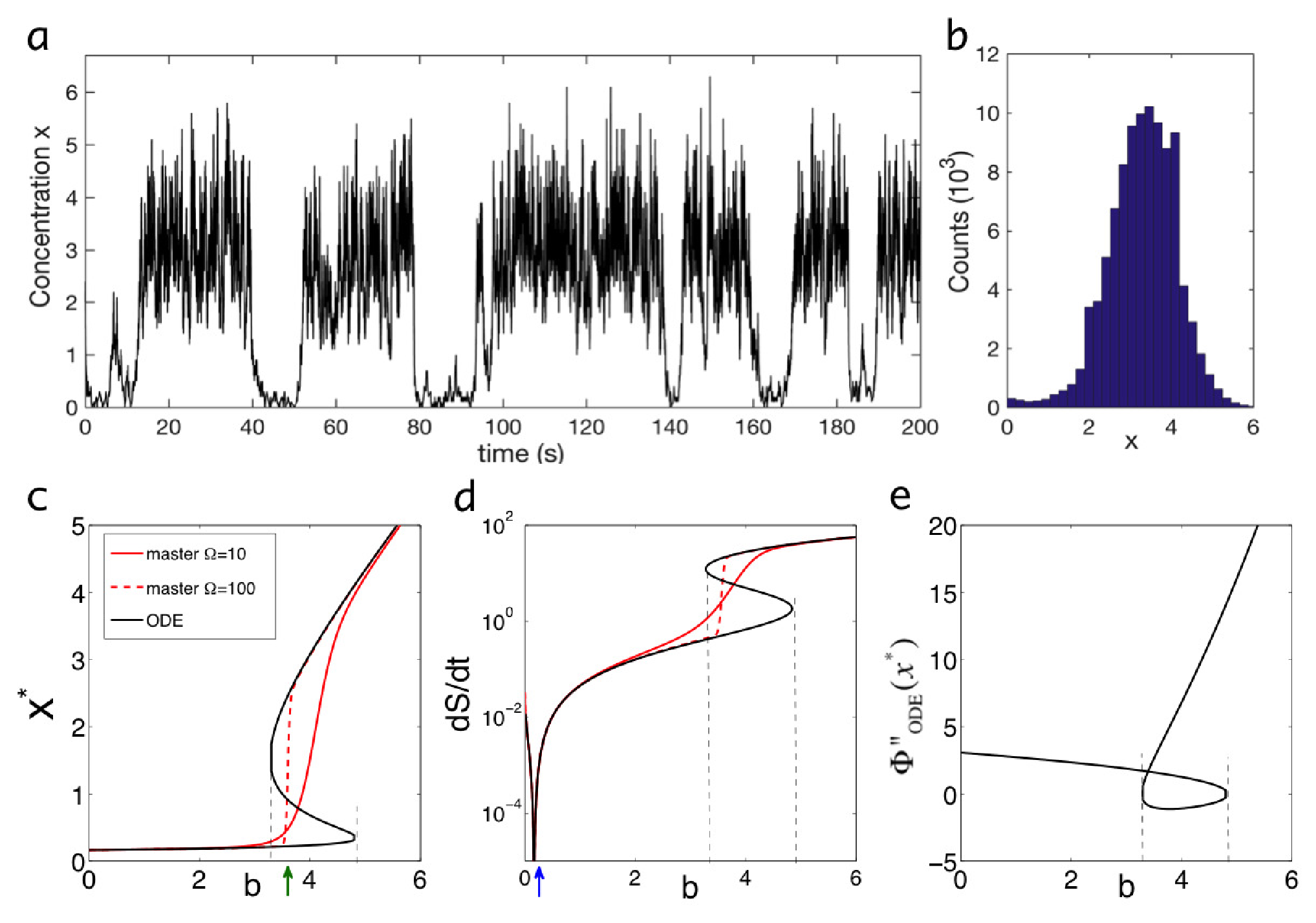}
\caption{{\bf Appearance of Keizer's paradox in Schl\"ogl model.} 
{(a) Gillespie simulations of Schl\"ogl model for $b=4$ and $\Omega=10$.
(b) Histogram of concentration levels for simulation from (a).}
(c) Bifurcation diagram (steady states) from ODE model (black). Average concentration from master equation for $\Omega=10$ (red solid line) 
and $\Omega=100$ (red dashed line). Quantity $b_c\approx 3.65$ represents the critical value (green arrow).
(d) Corresponding entropy production rates. Equilibrium $dS_i/dt=0$ occurs for $b_0=1/6$ (blue arrow).  
(e) Curvature (second spatial derivative) of effective potential from ODE model {evaluated at 
steady state}. Vertical grey dashed lines for guiding the eye. Parameters: $k_{+1}a=0.5, k_{-1}=3$, and $k_{+2}=k_{-2}=1$.}
\label{fig2}
\end{figure*}
\newpage
\ \\
\noindent{\bf Issues with previous attempts to disprove MaxEPP}\\
The Schl\"ogl model was used in the past to ‘disprove’ MaxEPP \cite{nicolis77} (reviewed in \cite{vellela09}). The argument goes 
as follows: In the ODE model, the high state always has the higher entropy production (see \textcolor{blue}{Fig. \ref{fig2}c,d}). 
This can easily be understood since overall in the Schl\"ogl model species $A$ is converted to species $B$ (and vice versa). 
Hence, $ds_i/dt = da/dt\cdot\Delta\mu/T\sim x$ with $\Delta\mu$ the chemical potential difference for the overall reaction, $T$ the 
temperature of the bath and $da/dt$ a linear function of $x$ (see Eq. \ref{A}). As a result, if MinEPP is the rule, 
then the low state should be selected, while MaxEPP would dictate that the high state is more stable. 

This argument can be made sharper when we consider the results from the master equation in the large volume limit. The average 
concentration $x$, as we discussed before, switches from the low to the high state at a critical value $b_c\sim 3.65$, with 
the switch becoming progressively sharper with increasing volume \cite{ge09,endres15}. This indicates a first-order phase transition and loss of bistability 
(\textcolor{blue}{Fig. \ref{fig2}c}, red curves). Hence, for $b<b_c$ the low state is selected, while for $b>b_c$ the high state is 
selected. Since these correspond respectively to the low and high entropy-production rates (\textcolor{blue}{Fig. \ref{fig2}d}, 
black curve), MinEPP (MaxEPP) would apply below (above) $b_c$, {and hence neither extremal principle would apply 
throughout.}

What is the issue with this conclusion? Because this and other models with large fluctuations suffer from 
Keizer’s paradox \cite{kurtz71,kurtz72,vellela07}, which says that microscopic (master equation) and macroscopic (ODE) descriptions can yield 
very different results, in particular when fluctuations play an important role (bistable systems, systems with 
possibility of extinction etc.). Ultimately, this paradox is caused by the switching of the order of the limits \cite{vellela09}. 
In the macroscopic description, the infinite volume limit is taken first (to derive the ODE), and 
then the infinite time limit is taken (for obtaining the steady states), while in the microscopic description 
the opposite order is applied. {Since the above argument combines the entropy production from the
bistable ODE model with the weights from the master equation, which is mono-stable in the infinite volume limit, 
this mixing of models may have led to the wrong conclusion regarding MaxEPP.}\\

\begin{figure}[t]
\includegraphics[width=8.5cm]{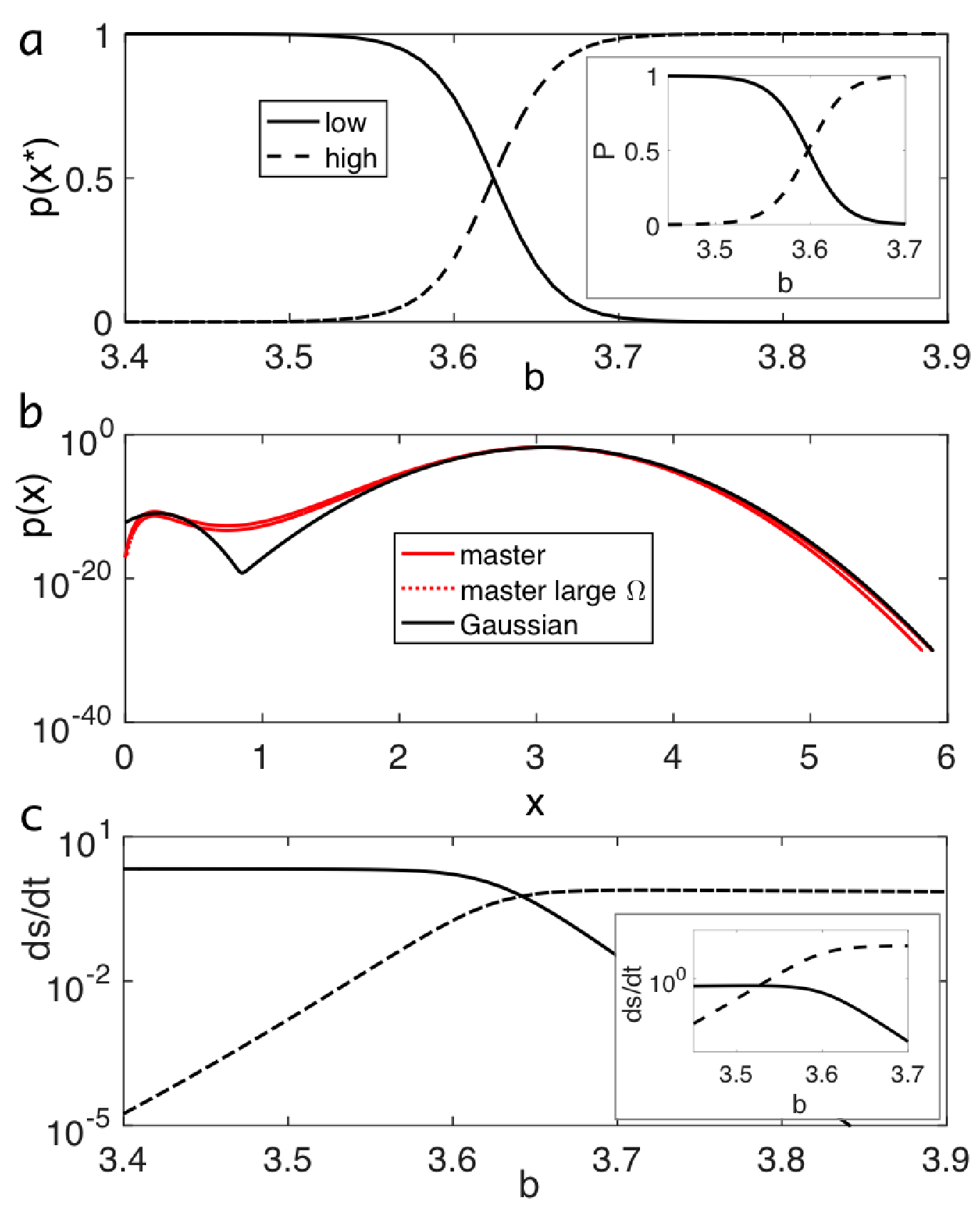}
\caption{{\bf MaxEPP in Schl\"ogl model.}
(a) Values of $p(x^*)$ from large-$\Omega$ limit of master equation evaluated at the low (solid curve)
and high (dashed curve) steady states $x^*$ for different $b$ values. 
{(inset) Weight of states from exact master equation by summing up probabilities for each peak in probability 
distribution (local minimum between low and high state is separatrix).}
(b) Distribution $p(x)$ from the exact master equation (red curve), large-$\Omega$ limit of the master 
equation (red dotted curve), and Gaussian approximation (black curve) for low and high states. 
(c) Entropy production rates calculated for each Gaussian peak. {(inset) Entropy productions from exact master
equation for each state by summing up contributions for each peak in probability distribution (local minimum between low 
and high state is separatrix).} (a-c) $\Omega=100$. Remaining parameters as in Fig. \ref{fig3}.}
\label{fig3}
\end{figure}

\noindent{\bf MaxEPP in a simple bistable model}\\
Having identified an inconsistency in the argument to disprove MaxEPP in the Schl\"ogl model, we now proceed to rescue MaxEPP as a valid
principle for determining the weights of steady states.
Specifically, we would first like to demonstrate that the order of the weights of the two stable steady states (\textcolor{blue}{Fig. \ref{fig3}a})
matches the order of their entropy production rates so that the state with the larger weight also has the larger entropy production rate. 
This would confirm the MaxEPP in this particular case. To obtain these rates, {we Taylor-expanded the exact potentials around the
steady-state values to obtain two Gaussian peaks for the weights of the two states} for $b$ near $b_c$ and large $\Omega$ 
(so that the analytical solution of the master equation is valid)
\begin{equation}
p_{k}(x)=Np(x_k^*)\sqrt{\frac{\Omega\Phi''(x_k^*)}{2\pi}}e^{\frac{-\Omega\Phi''(x_k^*)(x-x_k^*)^2}{2}}\label{gauss}
\end{equation}
with $k=1,2$ for the two Gaussian peaks centered around steady-state values $x_k^*$, the inverse of $\Phi''(x)=d^2\Phi(x)/dx^2$
proportional to the width of the peak, and $N=[p(x_1^*)+p(x_2^*)]^{-1}$ a normalization factor. 
Indeed, \textcolor{blue}{Fig. \ref{fig3}b} shows that the exact numerical solution of the master equation, the analytical large-$\Omega$ limit, 
and the {Gaussian approximation match reasonably well}. Now, using the expression for the average stochastic entropy production rate at 
steady state (Eq. \ref{Sedot2}) \cite{schnakenberg76,lebowitz99,gaspard04}
\begin{eqnarray}
\frac{dS_i}{dt}\!\!\!&=&\sum_{X,r}^{\pm2} W_{-r}(X|X\!+\!\Delta X_r)P(X\!+\!\Delta X_r)
\ln\frac{W_{-r}(X|X\!+\!\Delta X_r)}{W_{r}(X\!+\!\Delta X_r|X)}\nonumber\\
&\rightarrow&\Omega\sum_{r}\int dx\,w_{-r}(x)\,p(x)\ln\frac{w_{-r}(x)}{w_{r}(x)},
\end{eqnarray}
where the second line is valid for large $\Omega$, 
shows that the order of the rates indeed matches the order of the weights (\textcolor{blue}{Fig. \ref{fig3}c}). This can
be easily understood by approximating the $k^\text{th}$ peak by $\delta$-function $p_k(x)=Np(x_k^*)\delta(x-x_k^*)$,
resulting in $dS_{i,k}/dt=Np(x_k^*)\Omega\sum_r^{\pm 2}w_{-r}(x_k^*)\ln[w_{-r}(x_k^*)/w_{r}(x_k^*)]=Np(x_k^*)\Omega ds_{i,k}/dt$ with
$ds_{i,k}/dt$ the rate of macroscopic entropy production for state $k$ from Eq. \ref{sidot}. Hence, when
the weights of the states change from $p(x_1^*)>\!\!>p(x_2^*)$ to $p(x_1^*)<\!\!<p(x_2^*)$ for increased
driving, then $dS_{i,1}/dt>dS_{i,2}/dt$ changes to $dS_{i,2}/dt>dS_{i,1}/dt$  (even though $ds_{i,1}/dt < ds_{i,2}/dt$ 
always applies after dividing by $\Omega$, see \textcolor{blue}{Fig. \ref{fig2}d}). Consequently, the entropy production rate 
$dS_{i,k}/dt$, when correctly written as an extensive variable, can indeed be a proxy for the weight of a state $p(x_k^*)$.

Even without the Gaussian approximation, we can obtain the weights of the states and their entropy production rates
using the master equation. We can split up the contributing $x$ values into two parts (corresponding to the two states)
by using the separatrix as the natural attractor boundary (local minimum $x_{\rm min}\approx 0.9$ of probability distribution 
in \textcolor{blue}{Fig. \ref{fig3}b}). This produces the same qualitative result that the entropy production rates and weights of 
the two states are correlated (cf. insets of \textcolor{blue}{Fig. \ref{fig3}a,c}). Nevertheless, there are quantitative differences between 
the two approaches as the curves of the weights and the entropy production rates do not cross exactly at the same $b$ value.
This is because the order of the weights shown in \textcolor{blue}{Fig. \ref{fig3}a} also depends on the curvature $\Phi''(x)$ of the potential
in a nontrivial way. Consider the ratio of the transition rates between low (1) and high (2) states, given by \cite{vellela09,endres15}
\begin{equation}
\frac{r_{1\rightarrow 2}}{r_{2\rightarrow 1}}\sim \sqrt{\frac{\Phi''(x_1)}{\Phi''(x_2)}}e^{-\Omega[\Phi(x_1)-\Phi(x_2)]}\label{r_ratio}
\end{equation}
(see \textcolor{blue}{Supplementary Information} for details). The prefactor alone suggests that the lower the curvature the
higher the weight of a state but this weak curvature dependence (outside exponential) only reflects the attempt frequency 
to escape the attractor. In contrast,  \textcolor{blue}{Fig. \ref{fig2}e} suggests that the crossing of the curvature 
marks the transition (cf. \textcolor{blue}{Fig. \ref{fig2}c}). In particular, the more stable state (i.e. the low state below $b_c$
and the high state above $b_c$) appears to have the higher curvature. A higher curvature may imply a larger depth of the potential 
$\Phi_{\rm ODE}(x_k)$ and hence increased stability. However, a proper treatment requires the inclusion of noise, which is done
next.\\
\ \\
\newpage
\noindent{\bf General MaxEPP for nonequilibrium steady states}\\
Can we establish a formal link between the weight of a state and its entropy production and curvature in general? 
In the following, we approach the problem using the {Jaynes'} maximum caliber method \cite{jaynes80,presse13}, 
{which, put simply, is just an inference method similar to maximum entropy methods for equilibrium systems \cite{dewar09}.}
Basically, we wish to find the probability of a certain configuration for a system in a way that neither assumes something we 
do not know, nor contradicts something we do know.
For this purpose, we define the {caliber for the probability $p_\Gamma$ of observing a trajectory of duration $t$
\begin{equation}
C[P_\Gamma]=-\sum_{\Gamma}P_\Gamma\ln P_{\Gamma}-\lambda\sum_\Gamma P_\Gamma A_\Gamma,
\end{equation}
where the first term on the right-hand side is the Shannon information entropy and the second term is a constraint.} Our constraint 
is designed to implement that the action $A_\Gamma=\int_{0}^{t}L(\tilde t) d\tilde t$ is minimized with $\lambda$ a (positive) Lagrange multiplyer 
(reflecting our expectation that the observed average action and hence average difference in kinetic and potential energy are finite) 
and $L(t)$ the Lagrangian. Indeed, maximizing the entropy with respect to the probability of observing a trajectory
\begin{equation}
\frac{\delta C}{\delta P_\Gamma}=0\ \rightarrow\  P_\Gamma=\frac{e^{-A_\Gamma}}{Q}
\end{equation}
leads to a Boltzmann-like probability distribution {(with action in units of $\lambda^{-1}$)}, with $P_\Gamma$ the larger the smaller the action, 
representing the stochastic least-action principle \cite{wang06,seifert12}. Now, using the Evans-Searles fluctuation theorem \cite{kurchan98,lebowitz99}
\begin{equation}
\frac{P_\Gamma}{P_{-\Gamma}}=e^{\Delta S_\Gamma},\label{pgamma}
\end{equation}
where the ratio of the probabilities of forward and backward (time-reversed) trajectories corresponds to the 
exponential of the entropy produced along trajectory $\Gamma$ at steady state. Hence, the entropy production
\begin{equation}
\Delta S_\Gamma=\ln\frac{P_\Gamma}{P_{-\Gamma}}=A_{-\Gamma}-A_\Gamma\label{sig}
\end{equation}
is the difference between the backward and forward actions at steady state (see also \cite{seifert12}). 
While $A_\Gamma$ is minimal by construction, we have no information about $A_{-\Gamma}$ (MaxEPP would be proven if 
$A_{-\Gamma}$ is maximal). To gain insight into the problem we derive in the following the entropy production for 
steady states explicitly.

{To combine the best of ODEs and master equations, we extend Eqs. \ref{ODEx}-\ref{ODEb} by the following set of Langevin equations 
(stochastic differential equations)
\begin{subequations}
\begin{eqnarray}
\frac{dx}{dt}&=&\underbrace{w_{+1}-w_{-1}+w_{-2}-w_{+2}}_{-\Phi'_{\rm ODE}(x)}+\eta_x(t)\label{SDEx}\\
\frac{da}{dt}&=&\underbrace{-w_{+1}+w_{-1}}_{-\Phi'_{\rm ODE}(a)}+F+\eta_a(t)\label{SDEa}\\
\frac{db}{dt}&=&\underbrace{-w_{-2}+w_{+2}}_{-\Phi'_{\rm ODE}(b)}-F+\eta_b(t)\label{SDEb},
\end{eqnarray}
\end{subequations}
with short notations $\Phi'_{\rm ODE}(x)=\partial \Phi_{\rm ODE}(a,b,x)/\partial x$, 
$\Phi'_{\rm ODE}(a)=\partial \Phi_{\rm ODE}(a,b,x)/\partial a$ and $\Phi'_{\rm ODE}(b)=\partial \Phi_{\rm ODE}(a,b,x)/\partial b$ 
(although a unique $\Phi_{\rm ODE}(a,b,x)$ may not exist \cite{zhou12}). In Eqs. \ref{SDEx}-\ref{SDEb}, the $\eta$'s represent
noise terms. In the presence of noise, the steady state is now defined by the vanishing time derivatives of the averages 
$d\langle x\rangle^*/dt=d\langle a\rangle^*/dt=d\langle b\rangle^*/dt=0$, where $\langle a\rangle^*=a^*$,
$\langle b\rangle^*=b^*$, and $\langle x\rangle^*=x^*$.} 

There are a number of different ways of how to model the noise. In chemical reactions, noise is attributed to 
independent stochastic birth and death processes of the chemical species \cite{vanKampen}, leading to multiplicative noise 
of the form $\eta_a(t)=\sqrt{\epsilon}g_a\xi_a(t)$ and $\eta_b(t)=\sqrt{\epsilon}g_b\xi_b(t)$ with small parameter $\epsilon=1/\Omega$ and 
effective temperatures $g_a=g_a(a,b,x)=w_{+1}+w_{-1}$ and  $g_b=g_b(a,b,x)=w_{+2}+w_{-2}$ (and  $g_x=g_a+g_b$ for completeness). 
The fluctuations themselves ($\xi$'s) are 
assumed to be `white' with correlations $\langle\xi_k(t)\xi_{k'}(t')\rangle=\delta_{k,k'}\delta(t-t')$ and $k=a,b$. The noise in $X$ 
is just a reflection of the noises in $A$ and $B$ with $\eta_x(t)=-\sqrt{\epsilon}[g_a\xi_a(t)+g_b\xi_b(t)]$, thus avoiding double 
counting of noise contributions. For simplicity, we assume additive noise from now on, characterized by constant effective temperatures 
$g_k^*=g_k(a^*,b^*,x^*)$ as evaluated at the steady state. The usage of a constant effective temperature is a suitable approximation 
when the noise (or $\epsilon$) is small and the system is settled into its steady state (rare switching only; however, these effective 
temperatures can be different for the different states in a multistable system) \cite{hasty00}.

To reconnect with Eq. \ref{sig}, the stochastic action $A_\Gamma=A[{\mathbf q}(t)]$ of the combined dynamics of 
${\mathbf q}(t)=\{x(t),a(t),b(t)\}$ is required, {where we now introduce short notations 
$\dot {\mathbf q}=\{\dot x,\dot a, \dot b\}=\{dx/dt, da/dt, db/dt\}$ for the time derivatives.
According to \cite{arnold99,navarra13} the action for the system of Langevin equations \ref{SDEx}-\ref{SDEb} with additive noise is
\begin{widetext}
\begin{eqnarray}
A_\Gamma&=&\int_0^td\tilde t\left\{\frac{\Omega}{2}\sum_{q=a,b,x}\frac{[\dot q + \Phi'_{\rm ODE}(q)
-F\sigma_{q}]^2}{(g_{q}^*)^2}-
\frac{1}{2}\sum_{q=a,b,x}\Phi''_{\rm ODE}(q)\right\}\nonumber\\
&=&\int_0^td\tilde t\left\{\Omega\sum_{q=a,b,x}\underbrace{\left[\frac{\dot q^2 + [{\Phi'}_{\rm ODE}(q)-F\sigma_{q}]^2}{2(g_q^*)^2}\right.}_{E_{\rm kin}-E_{\rm pot}}
+\underbrace{\left.\frac{\dot q\Phi'_{\rm ODE}(q)}{(g_q^*)^2}\right]}_{\rm -(entropy\ prod.)}
-\underbrace{\frac{1}{2}\sum_{q=a,b,x}\Phi''_{\rm ODE}(q)}_{\rm noise}\right\},\label{A2}
\end{eqnarray}
\end{widetext}
with $\sigma_{a}=1$, $\sigma_b=-1$, and $\sigma_x=0$ for a trajectory of duration $t$ (related expressions can be derived for multiplicative noise, 
see \cite{haenggi89,wio89,zinn-justin96,tang14}). Note that the integral (or time average) over $F\dot q$ is zero and hence this term does not appear 
in Eq. \ref{A2}.} Furthermore, Eq. \ref{A2} shows that the curvature $\Phi''_{\rm ODE}$ of the potential does affect the probability of a trajectory, 
reflecting the disfavoring of high noise \cite{seifert12}. Now, combining Eqs. \ref{sig} and \ref{A2}, the entropy production is given by 
\begin{equation}
\Delta S_\Gamma=A_{-\Gamma} - A_\Gamma =-2\Omega t\left\{\frac{\overline{[\dot a\Phi'_{\rm ODE}(a)}]}{(g_a^*)^2}
+\frac{\overline{[\dot b\Phi'_{\rm ODE}(b)}]}{(g_b^*)^2}\right\}
\end{equation}
with $\overline{[...]}$ indicating time averaging for duration $t$ \cite{hatano01,seifert05}. Two comments are
in order. First, for the backward action, all time derivatives need reversing in sign, such as for $\dot a$, $\dot b$ and
$\dot x$. (The change in sign of $d\tilde t$ in the integral is canceled by the change in order of integration.)
Second, the term $\overline{[\dot x\Phi'_{\rm ODE}(x)]}$ is zero as there is no net flux in $x$ (see \textcolor{blue}{Supplementary Information}) 
\cite{tome97,tome06}.
 
In summary, the action in Eq. \ref{A2} contains a classical part (i.e. difference of kinetic and potential energies),  
a dissipative part (entropy production), and a stochastic part (curvature). Hence, trajectories do not only minimize the 
classical and stochastic actions (equivalent to solving the dynamical equations) but also maximize the entropy production 
(due to negative sign in front of the entropy production term). Eq. \ref{A2} thus results in the MaxEPP for trajectories 
in a multistable dynamical system, and is the main result of this paper.\\

\ \\
\noindent{\bf Simplified MaxEPP for nonequilibrium steady states}\\
The entropy production appearing in the action can be brought to a more familiar form, at least heuristically.
The entropy production for reservoir species $A$ can be rewritten as
\begin{equation}
\overline{[\dot a\Phi'_{\rm ODE}(a)]}=\overline{[(F-\Phi'_{\rm ODE}(a)+\sqrt{\epsilon}g_a^*\xi_a)\Phi'_{\rm ODE}(a)]}=
F^2-\overline{[\Phi'_{\rm ODE}(a)^2]}+\sqrt{\epsilon}g_a^*\overline{[\xi_a\Phi'_{\rm ODE}(a)]},\label{aPhi}
\end{equation}
and accordingly for species $B$. We need to be careful with the last term of Eq. \ref{aPhi}. Discretizing the Langevin equation, we obtain 
\begin{equation}
\sqrt{\epsilon}g_a^*\overline{[\xi_a\Phi'_{\rm ODE}]}=\frac{1}{2}\epsilon (g_a^*)^2\langle\Phi''_{\rm ODE}(a)\rangle\label{transform}
\end{equation}
(see \textcolor{blue}{Supplementary Information}) \cite{tome97,tome06,xiao09}, where we replaced time averages 
$\overline{[...]}=1/t\int_0^tdt...$ by ensemble averages $\langle\dots\rangle=\int da\, p(a,b,x)...$ at steady state, valid for 
sufficiently long trajectories. (However, this ensemble average is technically restricted to sampling from a particular steady state, 
as we do consider switching between states here.)  
According to Eq. \ref{transform} the last term in Eq. \ref{aPhi} is smaller than the first two by a factor $\epsilon$. 
Now, introducing ensemble averages throughout, we can, at least heuristically, introduce the potential from the master equation
\begin{equation}
\frac{\langle{\Phi'_{\rm ODE}(a)}^2\rangle}{(g_a^*)^2}
\approx\langle\Phi'_{\rm ODE}(a)\Phi'(a)\rangle=\left\langle\Phi'_{\rm ODE}(a)\ln\left(\frac{w_{+1}}{w_{-1}}\right)\right\rangle
\end{equation}
using the detour of the Fokker-Planck potential as another approximation to the master equation (see \textcolor{blue}{Supplementary Information} for details).
Next, using Jensen's inequality for convex functions, we obtain the approximate steady-state weight
$p(x^*)\approx \langle p_\Gamma\rangle_{x^*}=\langle e^{-A_\Gamma}\rangle/Q\geq e^{\langle A_\Gamma\rangle}/Q$,
which becomes an exact equality for small noise. Keeping only the highest order terms in $\Omega$ (or lowest order terms in $\epsilon$) 
and using Eq. \ref{A2}, the steady-state weight becomes
\begin{widetext}
\begin{equation}
p(x^*)\sim\exp\left\{-\Omega t\left[\sum_{q=a,b,x}
\underbrace{\frac{\langle\dot q^2\rangle + \langle[{\Phi'}_{\rm ODE}(q)-F\sigma_{q}]^2\rangle}{2(g_q^*)^2}}_{\rm class.\ action\ (min.)}
\ -\ \frac{1}{2}\sum_{r=1,2}\underbrace{\left\langle(w_{+r}-w_{-r})\ln\left(\frac{w_{+r}}{w_{-r}}\right)\right\rangle}_{\rm entropy\ prod.\ (max.)}
\right]\right\},\label{A3}
\end{equation}
\end{widetext}
up to normalization and valid for durations $t$ much smaller than the time scale for switching states. 
In Eq. \ref{A3}, the entropy production has now the familiar form given by Eq. \ref{sidot} (also known as 
Schnakenber's formula \cite{schnakenberg76}). Typically, at steady state we have
$\langle\dot a^2\rangle\approx\langle\dot b^2\rangle\approx\langle\dot x^2\rangle\approx\langle\Phi'_{\rm ODE}(x)^2\rangle\approx 0$, as well as 
$\langle\Phi'_{\rm ODE}(a)^2\rangle\approx-\langle\Phi'_{\rm ODE}(b)^2\rangle\approx F$, although tradeoffs among the different terms can occur.
Hence, the main difference between the different steady states in a multistable system is the entropy production term, which depends on
the steady-state value $x^*$.

{Taken together, Eq. \ref{A3} demonstrates once more the roles of {\it both} the classical and the dissipative action in 
determining the probability of a steady state.  Hence, MaxEPP is a principle for multistable systems 
in which the entropy production biases the evolution of the system towards the highest-entropy producing state.}\\

\begin{figure}[b]
\includegraphics[width=9cm]{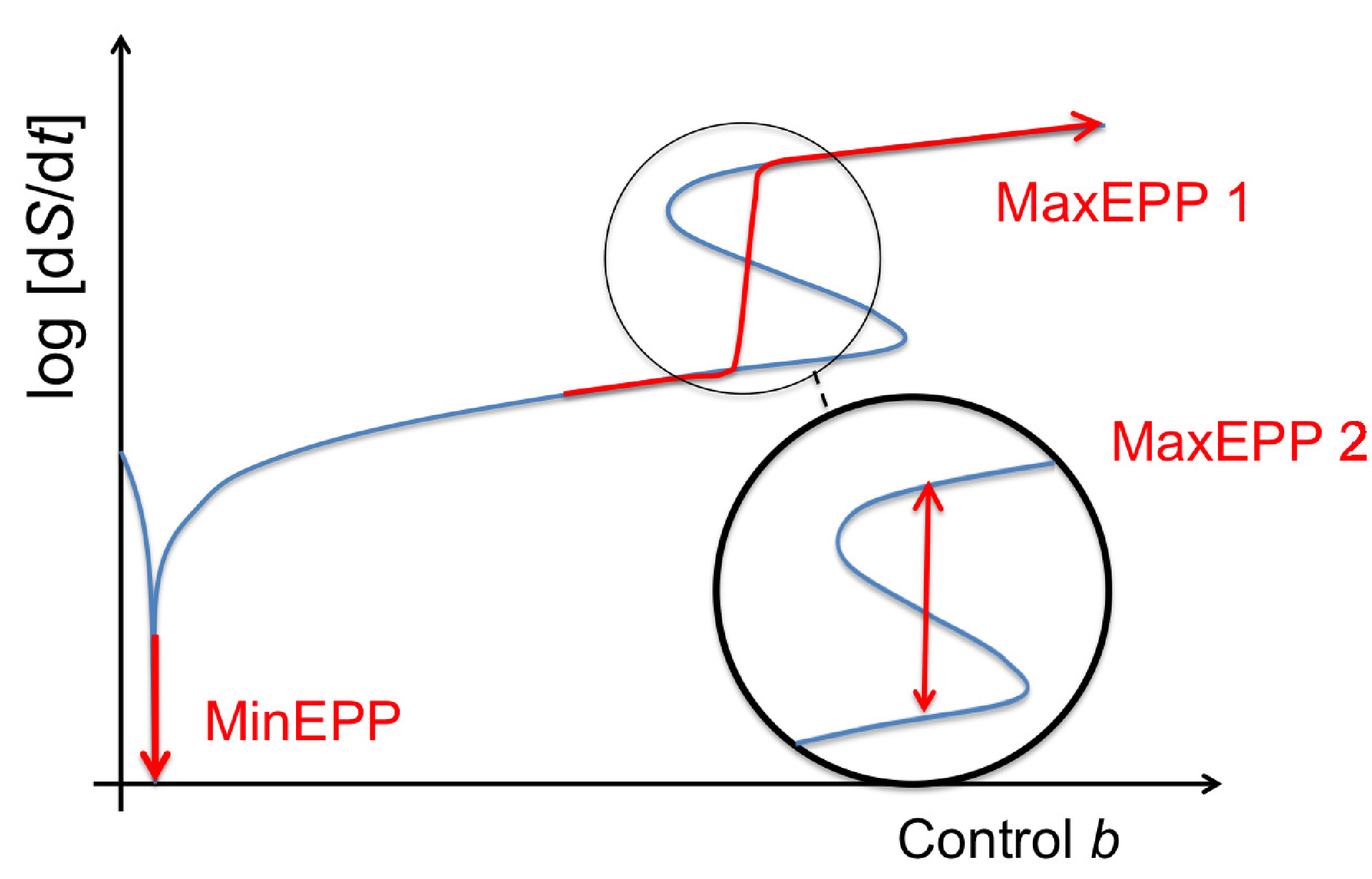}
\caption{{\bf Min- and MaxEPPs.}
Illustration of MinEPP and two different MaxEPP in a semi-log plot of entropy production rate $dS/dt$ versus control parameter b. 
MinEPP is valid near equilibrium, where $dS/dt\gtrsim 0$. MaxEPP 1 simply states that the more a system is driven away
from equilibrium the more entropy is produced. MaxEPP 2 is more subtle, describing how states are selected in a multistable
system.} 
\label{fig4}
\end{figure}

\noindent{\bf Discussion}\\
We showed that MaxEPP is applicable when comparing the two states in the simple bistable Schl\"ogl model (using the master equation;
\textcolor{blue}{Fig. \ref{fig3}a,c}) 
and when considering trajectories of a multistable system at steady state in the large-volume limit (using the Langevin approximation;
Eq. \ref{A3}). 
MaxEPP applies in the former because the weights of the low and high states shift in the exact stochastic approach 
due to a first-order phase transition. MaxEPP applies in the latter because trajectories minimize the action two-fold: 
First, the classical action is minimized, meaning that the dynamic system takes on its appropriate solution, i.e. $\dot x=-\Phi'_{\rm ODE}(x)$.
Second, the entropy production from the fluxes between the reservoir and the reaction volume is maximized. Hence, in a multistable system, the steady state with 
the highest entropy production is naturally selected {(similar to the `state selection' principle proposed
in \cite{meysman10})}. Our analytical derivations show that MaxEPP is a consequence of the least-action principle 
applied to dissipative systems (stochastic least-action principle). Note however the discrepancy in how the MaxEPP is achieved in the 
two approaches: using the master equation we observe a first-order phase transition and state switching at the critical point, while using the
Langevin approximation, the high state is selected.

In addition to this local MaxEPP for states of a multistable system for a {\it fixed} driving force, there is also 
a trivial global MaxEPP principle, which simply says that the more a system is driven away from equilibrium the 
more it produces entropy {(`gradient response' principle) \cite{schneider94,meysman10}.} This statement is simply a result 
of the average of Eq. \ref{FT} (or Eq. \ref{sig}) given by the Kullback-Leibler divergence between the forward and backward trajectory 
distributions $P_\Gamma$ and $P_{-\Gamma}$, respectively, 
\begin{equation}
\langle \Delta S_\Gamma\rangle=\sum_\Gamma P_\Gamma\ln\left(\frac{P_\Gamma}{P_{-\Gamma}}\right)\geq 0,\label{sig_ave}
\end{equation}
which has been mentioned before \cite{searles08}. Eq. \ref{sig_ave} is minimally zero (at equilibrium due to detailed balance) 
and is the larger the more the forward and backward trajectories differ. Hence, the earlier discussed MinEPP is not really a separate
principle, but simply a different perspective of the global MaxEPP. \textcolor{blue}{Fig. \ref{fig4}} summarizes the two MaxEPPs 
and the MinEPP for a bistable system at steady state. 

Our results can be connected to recent results in fluid systems. Similar to the Schl\"ogl
model with a nonequilibrium first-order phase transition, flow systems undergo a  
laminar-turbulent flow transition as the Reynolds number ($Re$) increases \cite{martyushev07}. In both systems, MaxEPP applies and can be
used to predict the critical transition point (\textcolor{blue}{Fig. \ref{fig3}a,c} in the former and fig. 2 of \cite{martyushev07} in the latter). 
What are the weights of the states in the fluid system? As a macroscopic system, the system is largely monostable - below the 
critical $Re$, laminar flow is dominant, while above it, turbulent flow is the result. This is analogous to the
macroscopic Schl\"ogl model, where the bistable region disappears for increased system size and a first-order phase transition results 
(\textcolor{blue}{Fig. \ref{fig2}c}). 
However, even in the fluid system, the laminar state can be metastable even for relatively large $Re$ if unperturbed. This is a sign
of hysteresis and hence bistability, and so both laminar and turbulent flows may coexist with the turbulent flow the more
stable state (turbulent flow never switches back to laminar flow when $Re$ is above the critical value).

Another previously investigated physical system is the fusion plasma, where a thin layer of fluid is heated 
from one side. Models of heat transport in the boundary layer predict that the MaxEPP (MinEPP) applies when the heat flux 
(temperature gradient) is fixed \cite{yoshida08,kawazura10}. 
Similarly in the Schl\"ogl model, when the concentrations of species A and B are fixed (like the temperature in the fusion plasma), 
the system becomes indistinguishable from an equilibrium system (see comments
above Eq. \ref{ODEx}) and the entropy production is zero (extreme version of MinEPP). However, once fluxes are fixed 
(Eqs. \ref{ODEa} and \ref{ODEb}, or Eqs. \ref{SDEa} and \ref{SDEb}), the MaxEPP results (Eq. \ref{A2}). We believe that flux constraints correspond 
to the more physically correct scenario as now the entropy production of the macroscopic (ODE) model matches the entropy production of the microscopic model as 
described by the exact master equation (cf. Eqs. \ref{Sedot} and \ref{sidot}). 
In both above described flow systems, dissipative structures form when strongly driven. In the former fluid system, turbulent swirl structures appear 
while in the latter plasma system a shear flow is induced. What do such dissipative structures correspond to in the Schl\"ogl model? 
There are large fluctuations and inhomogeneities in the spatial Schl\"ogl model with diffusion for increasing system size, 
although these may represent the approach of the critical point and less actual dissipative structures \cite{endres15}.

Paradoxically, work in the fluid system raised the possibility that both MinEPP and MaxEPP apply simultaneously. 
MinEPP appears to predict the flow rates in parallel pipes while MaxEPP seems to predict the flow regime (laminar versus turbulent) \cite{niven10}. 
This can potentially be explained by our Eq. \ref{A3} as follows:
The first term, which represents the classical action, may lead to a reduced entropy production (and potentially MinEPP), if the 
effective temperature $g_x^*$ (noise) is small. In this case, the entropy production (second term) is less crucial to fulfill. 
In contrast, if the noise is large, the classical action becomes a less important constraint, and the entropy production becomes 
important, leading necessarily to MaxEPP.

The MaxEPPs was previously also applied to ecosystem functioning, which aims to predict the evolution of 
large-scale living systems in terms of thermodynamics (also called ecological thermodynamics). Considering simple food-web 
models of predators, preys, and other resources, the state-selection and gradient-response principles were found to break down in more 
complicated models with multiple trophic (hierarchical) levels \cite{meysman10}. However, the stability of the steady states was 
assessed with linear stability analysis, i.e. through the response to small perturbations around the macroscopic steady state.
However, as we showed, the macroscopic Schl\"ogl model predicts the wrong stability and only in the thermodynamic limit of the
microscopic master-equation model the MaxEPP is predicted correctly.

Our interpretation of MaxEPP is in line with the recent finding that the entropy production, by itself, 
is not a unique descriptor of the steady-state probability distribution \cite{zia06}. According to Eq. 
\ref{A2}, other terms matter for the probability of a trajectory, such as the classical action and terms disappearing in the limit 
of large $\Omega$. In fact, far-from-equilibrium physics has many pitfalls. While Eq. \ref{sig_ave} leads to safe predictions 
about the expected entropy production, the fluctuation theorem for individual trajectories given by Eq. \ref{pgamma}, 
i.e. $P_\Gamma/P_{-\Gamma}=\exp(\Delta S_\Gamma)$, has to be treated with caution. A trajectory $\Gamma$ with a large ratio 
$P_\Gamma/P_{-\Gamma}$ is not necessarily selected because it has a large entropy production; $P_\Gamma$ might still be tiny 
(and $P_{-\Gamma}$ even tinier) so that $\Gamma$ is extremely unlikely to occur. Whether a trajectory is actually selected depends on 
the underlying chemical rules or physical laws (see classic action in Eq. \ref{A2} and \textcolor{blue}{Supplementary Information} 
with \textcolor{blue}{Fig. S3} for an explicit example).

Future work may investigate applications of MaxEPP in models of nonequilibrium self-assembly, climate, and the
emergence of molecular complexity (or life). Imagine there are two stable steady states, 
one with high complexity and high entropy production, and another one with low complexity and low entropy production. We 
speculate that the high complexity state is more likely as long as the extra cost from the entropy
reduction due to complexity is offset by a significantly larger entropy production. Another issue to keep in mind is that 
evolution of our biosphere may not be at steady state, and so transient behavior may need to be investigated.
\ \\ \ \\


\noindent{\bf Acknowledgements:} \\
R.G.E. thanks Audrey Yurika Marvin for help with the simulations, T\^ania Tom\'e for 
numerous helpful discussions, and Linus Schumacher for a critical reading of the manuscript. R.G.E. also thankfully acknowledges 
financial support from the European Research Council Starting Grant N. 280492-PPHPI and BBSRC grant BB/G000131/1.\\
\ \\
\noindent{\bf Author contributions:}\\
R.G.E. conceived the study, developed the models and simulations, and wrote the paper.\\
\ \\
\noindent{\bf Additional information:}\\
Supplementary information accompanies this paper at http://www.nature.com/scientificreports\\
\ \\
Competing financial interests: The author declares no competing financial interests.
\ \\ \ \\

\end{document}